\documentclass[sigconf,screen]{acmart}
\AtBeginDocument{%
  \providecommand\BibTeX{{%
    \normalfont B\kern-0.5em{\scshape i\kern-0.25em b}\kern-0.8em\TeX}}}

\copyrightyear{2023}
\acmYear{2023}
\setcopyright{rightsretained}
\acmConference[IUI '23]{28th International Conference on Intelligent User Interfaces}{March 27--31, 2023}{Sydney, NSW, Australia}
\acmBooktitle{28th International Conference on Intelligent User Interfaces (IUI '23), March 27--31, 2023, Sydney, NSW, Australia}
\acmDOI{10.1145/3581641.3584059}
\acmISBN{979-8-4007-0106-1/23/03}

\usepackage{microtype}
\usepackage{tikz}

\usepackage{paralist}
\usepackage{color,soul}
\usepackage{xspace}
\usepackage{xcolor}
\usepackage{multirow}
\usepackage{verbatim}
\usepackage{multicol}
\usepackage{enumitem}
\usepackage{etaremune}
\usepackage{soul}
\usepackage[normalem]{ulem}
\usepackage{subcaption,booktabs}
\usepackage{mathtools}
\usepackage{comment}
\usepackage{cleveref}
\usepackage{subcaption}
\usepackage{array}

\usepackage{pifont}% http://ctan.org/pkg/pifont

\newcommand{\cmark}{{\ding{51}}}%
\newcommand{\xmark}{{\ding{55}\ }}%
\definecolor{cxmark}{HTML}{EC6742}
\definecolor{ccmark}{HTML}{499D5E}

\newcommand{\cmarkcolor}{{\color{ccmark}{\cmark}}\xspace}%
\newcommand{\xmarkcolor}{{\color{cxmark}{\xmark}}\xspace}%

\definecolor{cprompt}{HTML}{3c4043}
\newcommand{\exinline}[1]{{\color{cprompt}``#1''\xspace}}
\newcommand{\quoteinline}[1]{{\color{cprompt}\emph{``#1''}\xspace}}

\newcommand{\fixed}[1]{{\color{blue}#1}}
\newcommand{\reviews}[1]{{\color{purple}#1}}

\renewcommand{\fixed}[1]{#1\xspace}
\renewcommand{\reviews}[1]{}

\newcommand{\eg}{\emph{e.g.,}\xspace}%
\newcommand{\ie}{\emph{i.e.,}\xspace}

\newcommand{\paragraphBold}[1]{\paragraph{\emph{\textbf{#1}}}}

\newcommand{\sysname}{\textsc{ScatterShot}\xspace}
\newcommand{\repourl}{https://github.com/tongshuangwu/scattershot}

\newcommand*\textcircle[1]{\tikz[baseline=(char.base)]{
            \node[shape=circle,draw,inner sep=0.5pt] (char) {#1};}}

\begin{document}

\title[Interactive In-context Example Annotation for Text Transformation]{\sysname: Interactive In-context Example Curation \\for Text Transformation}

\author{Tongshuang Wu}
\authornote{The work was mostly done when the first author was a PhD student at the University of Washington.}
\email{sherryw@cs.cmu.edu}
%\orcid{0000-0003-1630-0588}
\affiliation{%
  \institution{Carnegie Mellon University}
  \country{USA}
}

\author{Hua Shen}
\email{huashen218@psu.edu}
\affiliation{%
  \institution{Pennsylvania State University}
  \country{USA}
}

\author{Daniel S. Weld}
\email{weld@cs.uw.edu}
\affiliation{%
  \institution{University of Washington \&}
  \institution{Allen Institute for Artificial Intelligence}
  \country{}
}

\author{Jeffrey Heer}
\email{jheer@cs.uw.edu}
\affiliation{%
  \institution{University of Washington}
  \country{USA}
}

\author{Marco Tulio Ribeiro}
\email{marcotcr@microsoft.com}
\affiliation{%
  \institution{Microsoft Research}
  \country{USA}
}

%\end{comment}
%\renewcommand{\shortauthors}{G. Bansal et al.}


\begin{abstract}
The in-context learning capabilities of LLMs like GPT-3 allow annotators to customize an LLM to their specific tasks with a small number of examples.
However, users tend to include only the most obvious patterns when crafting examples, resulting in underspecified in-context functions that fall short on unseen cases. 
Further, it is hard to know when ``enough'' examples have been included even for known patterns.
% However, users tend examples tend to repeat a few obvious patterns, resulting in underspecified in-context functions that fall short on unseen cases. 
% Even if they are aware of some difficult patterns, it is unclear when they should consider a pattern learned. 
In this work, we present \sysname, an interactive system for building high-quality demonstration sets for in-context learning. \sysname iteratively slices unlabeled data into task-specific patterns, samples informative inputs from underexplored or not-yet-saturated slices in an active learning manner, and helps users label more efficiently with the help of an LLM and the current example set. %with the help of the current version of 
% In essence, \sysname enables active learning for in-context learning --- it iteratively identifies clusters of unlabeled inputs with similar task-specific patterns, samples inputs with difficult or underexplored patterns, and inquires human feedback on the current in-context function’s output on these samples.
% As humans evaluate and fix these samples, \sysname helps them iteratively reflect on the quality of the current function, and expand the in-context example set with informative demonstrations.
In simulation studies on two text perturbation scenarios, \sysname sampling improves the resulting few-shot functions by 4-5 percentage points over random sampling, with less variance as more examples are added.
In a user study, \sysname greatly helps users in covering different patterns in the input space and labeling in-context examples more efficiently, resulting in better in-context learning and less user effort.
%quality more stably than random sampling as the example set expands, with 4-5 percentage points improvement in final function performance.
% Through user studies, we also find that \sysname facilitates users' exploration of the input space, enables them to craft in-context examples more efficiently, and ultimately helps them build better in-context functions.
\end{abstract}

\begin{comment}
\begin{teaserfigure}
\centering
\includegraphics[trim={0cm 20cm 11cm 0cm}, clip,width=1\linewidth]{figures/teaser_temporal.pdf}
\vspace{-10pt}
\label{fig:teaser}
\end{teaserfigure}
\end{comment}


\maketitle

\section{Introduction}
\label{sec:intro}

\begin{figure*}
\centering
\includegraphics[trim={0cm 20cm 11cm 0cm}, clip,width=0.95\linewidth]{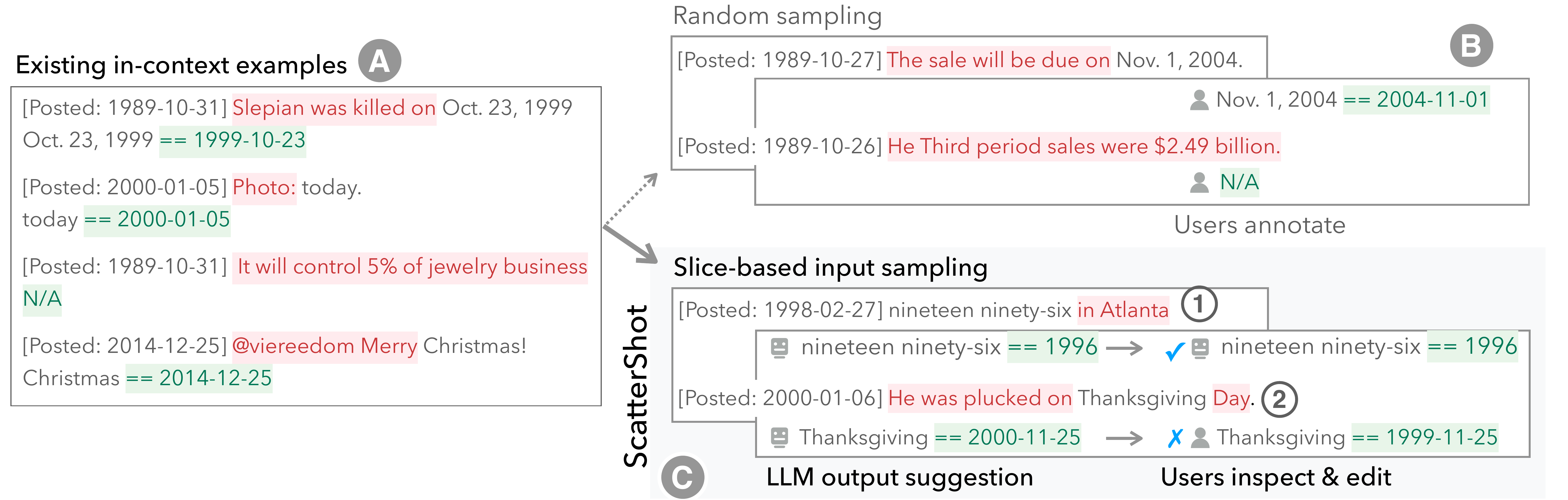}
\vspace{-5pt}
\caption{
An overview of how human annotators can use \sysname to iteratively collect effective in-context examples for temporal expression extraction and normalization.
The function extracts phrases with temporal meaning from sentences (\eg \exinline{Oct. 23, 1999} in \exinline{Slepian was killed on Oct. 23, 1999}), and normalizes them into standard formats (\exinline{Oct. 23, 1999 == 1999-10-23}) --- the red spans represent information deleted from the input, and the green ones represent information generated in the output.
Given an in-context example set that is likely \emph{underspecifying} the intended functionality (A), 
\sysname applies slice-based sampling to return unlabeled inputs that either have novel patterns or are difficult cases, and uses the existing examples to drive an LLM (\eg GPT-3) to suggest (possibly noisy) annotations, such that humans can correct the suggested annotations and possibly expand the in-context example bucket.
Compared to random sampling and manual labeling (B), \sysname helps humans re-allocate annotation budgets towards informative examples, and increases the in-context function performance.
}
\vspace{-5pt}
\label{fig:teaser}
\end{figure*}

% The emergence of large language models (LLMs) introduces new possibilities for developing customized functions (or completing customized tasks) without training models from scratch.
% Pre-trained on a large amount of text data, models like GPT-3~\cite{brown2020language} and Jurassic-1~\cite{J1WhitePaper} encode enough information to support \emph{in-context learning}~\cite{xie2021explanation}:
% they can be easily customized at run time (without any re-training needed) to handle new tasks, simply by taking in natural language instructions (\emph{prompts}) constructed with a few examples of the task at hand.
% For example, one could adapt an LLM to act as a translation engine, simply by prepending a few examples of the demonstrative input-output pairs, before the desired input:
% \exinline{Holiday: Christmas => Date: 12-25; Holiday: Halloween => Date: 10-31; Holiday: Thanksgiving =>}
% The LLM will then mimic the input pattern, and compute the most likely continuation --- the date for the input \exinline{Thanksgiving}, which could be \exinline{11-24}.
% Beyond this toy example, prompting has been used to achieve various ML functionalities in real-time, including code generation, question answering, creative writing, and others~\cite{swanson2021story, mishra2021natural, brown2020language}.

In-context learning \cite{xie2021explanation} is a property of Large Language Models (LLMs), where a user can ``write'' a transformation function via an (optional) short set of instructions and a few (input, output) examples.
For example, writing a function that ``translates'' a holiday name (e.g. ``Christmas'') into its calendar date (e.g. ``12/25'') would previously require a complicated rule-based system capable of mapping various kinds of subtly different inputs (e.g. \exinline{Xmas}, \exinline{Christmas day}, etc) to a lookup table of dates. 
With LLMs like GPT-3 \cite{brown2020language}, the process is much simpler.
A user can achieve the same functionality  with a \emph{prompt} (\ie a natural language instruction) that contains a small number (\eg two) of simple demonstrations, followed by a query (underlined):
\exinline{Christmas => 12/25; Halloween => 10/31; \uline{Independence Day (US)} =>}.
GPT-3 would take the prompt and return the right date \exinline{7/04} for this query. 
More impressively, LLM will also have some generalizability towards semantically relevant queries, \eg queries with abbreviations (\exinline{xmas => 12/25}, \exinline{nye => 12/31}), misspellings (\exinline{s patrics day => 03/17}), lesser-known name variations (\exinline{All Saints' Eve => 10/31}), and holidays that might be overlooked (\eg \exinline{Harriet Tubman Day => 3/10}).
The much reduced programming effort (compared to \eg rule-based systems) draws users' attention towards building their personalized in-context functions in various use scenarios, including code generation, question answering, creative writing, and others~\cite{swanson2021story, mishra2021natural, wu2022promptchainer}.

Although in-context learning has great potential, the quality of the learned function for non-trivial tasks depends on which in-context examples are used as demonstrations \cite{rubin2021learning, lu2021fantastically}.
Techniques for automatic example selection \cite{liu2021makes} depend on existing labeled datasets and tasks that can be evaluated automatically (\eg classification), and thus users ``in the wild'' rely on their own ingenuity and intuition when coming up with demonstrations~\cite{jiang2022prototype}.
%, leading to the new ``discipline'' of \emph{prompt engineering}
Unfortunately, users tend to focus on the most obvious and memorable patterns for demonstration \cite{gururangan2018annotation}, leading to systematic omissions \cite{wu-etal-2021-polyjuice} and \emph{underspecification} that might go unnoticed.
As an example, in Figure~\ref{fig:teaser} we use in-context learning to build a function to extract and normalize temporal information from a sentence \cite{chang2012sutime}.
Most users would provide demonstrations with common date formats (e.g. \exinline{Oct. 23, 1999}), and some might remember relative date references (e.g. \exinline {today}).
However, some patterns are easy to miss, e.g. long-form dates with no capitalization or holidays (e.g. \exinline{nineteen ninety-six}, \exinline{Thanksgiving Day} in Figure \ref{fig:teaser}C), and the LLM may fail to learn them if they are omitted.
Even sampling random examples from the unlabeled data might lead to the repetition of common patterns (Figure~\ref{fig:teaser}B) at the expense of demonstrating less-common ones.
What is worse, users may not know when they have provided enough examples, and whether there are any uncovered patterns in the unlabeled data.
As a result, prior work summarized the two major pain points of prompting to be (1) the effort required to source examples for a prompt, and (2) the difficulty of evaluating whether a prompt is improving~\cite{jiang2022promptmaker}.

In this work, we present \sysname, an interactive system for building high-quality demonstration sets for in-context learning. 
In a nutshell, \sysname helps users find informative input examples in the unlabeled data, annotate them efficiently with the help of the current version of the learned in-context function, and estimate the quality of said function.
In each iteration, \sysname automatically slices the unlabeled data into clusters based on task-specific \emph{key phrases} \cite{wu2020tempura, wu-etal-2021-polyjuice}.
For example, given the existing examples in Figure \ref{fig:teaser}A, it finds a cluster based on holiday key phrases (\exinline{Christmas}, \exinline{Thanksgiving}, etc.) and a cluster based on exact dates like \exinline{Oct. 23, 1999} (among others). 
\sysname keeps a running estimate of the error of each cluster, and thus prioritizes examples from clusters that have not yet been explored or learned effectively. It further uses the \emph{stability} of the current in-context function with respect to minor changes in the prompt (e.g. ordering of in-context examples), prioritizing unlabeled examples that get different predictions with different prompt variations. 
Users are then presented with examples of underexplored clusters (\eg Figure \ref{fig:teaser} C$_1$), or hard examples of explored clusters (\eg C$_2$, hard because the past tense refers to the Thanksgiving date of the \emph{previous} year).
Instead of having to label demonstrations from scratch, users can either accept correct predictions from the current function (Fig \ref{fig:teaser} C$_1$) or make edits to fix wrong predictions (Fig \ref{fig:teaser} C$_2$).
These additional labels are used to update the in-context function, such that the user explores the different possible input patterns in an interactive manner, without wasting resources on patterns that have already been learned.

\begin{figure*}
\centering
\includegraphics[trim={0cm 19.5cm 18cm 0cm}, clip,width=0.95\linewidth]{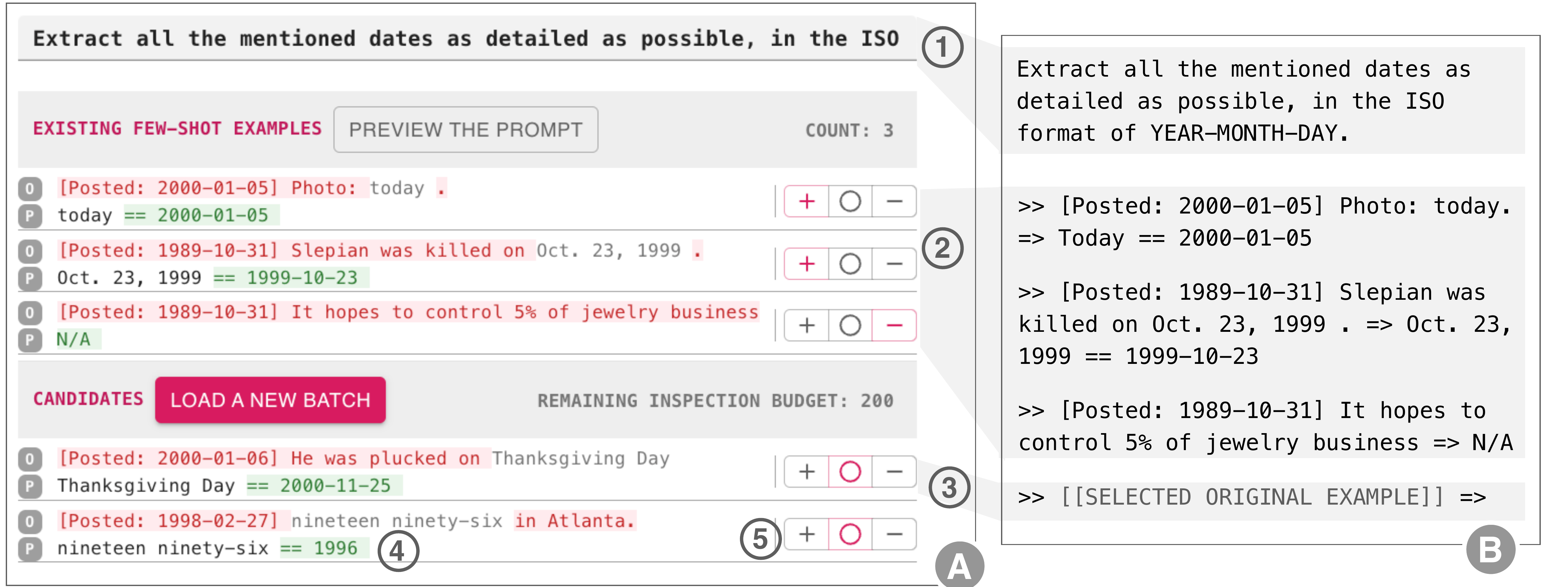}
\vspace{-5pt}
\caption{
(A) The \sysname interface, with (A$_1$) task description, (A$_2$) existing in-context examples, and (A$_3$) candidate examples awaiting human inspection. 
Through interactions A$_4$ and A$_5$.
Users can make edits to LLM outputs, sort the candidates into positive demonstrative examples (+), negative ones (-), or just not include the candidate (O).
The description and the examples get transformed into raw text prompts.
One set of in-context examples produces multiple prompts depending on how the examples are ordered; (B) shows a prompt with one possible ordering.
}
%\vspace{-5pt}
\label{fig:ui}
\end{figure*}

We evaluate \sysname both in terms of sampling efficiency and support for human annotators.
In simulation experiments, we compare the sampling strategy in \sysname{} to random sampling on two text transformation tasks contemplated in prior work: the data wrangling task illustrated in Figure \ref{fig:teaser}~\cite{chang2012sutime}, and rewriting question-answer pairs into logically equivalent pairs in order to evaluate model consistency \cite{ribeiro2019red}. 
In both cases, we find \sysname{} improves performance on corresponding metrics (\eg Rouge-L, F1) by 4-5 percentage points, with less variance for various values of $k$ demonstrations. 
Further, we conduct a within-subject user study in which 10 participants build in-context functions for the QA-pair rewriting task either (1) manually, (2) with the \sysname interface but random sampling, or (3) with the fully-featured \sysname.
We show that \sysname's interface alone is an improvement, by offloading input selection and providing sample outputs. 
Moreover, the sampling strategy in the fully-featured \sysname helps users notice diverse input patterns, leading to improvements in the resulting in-context function.
For example, participants who thought their in-context examples were sufficient when using random samples labeled \emph{an additional} 1.4 times of examples after switching to full \sysname (as they found new patterns), which further improved the function test performance.
We conclude the paper with insights into challenges and opportunities that arise from our experiments, including 
\eg explaining the sampling rationales, incorporating automated blind-spot detection, and the potential of using a \sysname setup to help users iteratively refine their task definition during data collection.

% \begin{itemize}
%     \item We introduce \sysname, an interactive system that helps annotators craft iteratively in-context examples, and raise the importance of effective input sampling, semi-automated labeling, and in-context function performance estimation.
%     \item We design and implement \emph{slice-based} sampling, which inquiries human annotators with examples selected from not-yet-learned slice, and shows the important of building \emph{task-dependent} data slices while eliminating irrelevant noises.
%     \item We report results from both a simulation study and a 10-person evaluation showing \sysname can increase in-context function performance, as well as annotator's awareness on diverse and edge cases. Our findings point to interesting future directions, including \eg explain the sampling rationales, incorporating automated blind-spot detection, and the potential of using a \sysname setup to help annotators iteratively refine their task definition during data collection.
% \end{itemize}

%\section{Design and Implement \sysname}
\section{The Design of \sysname}
The goal of \sysname is to help users iteratively find and label high-quality demonstrative examples to build effective in-context functions. In order to be effective, a function must be able to handle \textbf{common patterns} (\eg the temporal normalization function in Figure \ref{fig:teaser} must be able to handle common temporal expressions such as \exinline{today}), \textbf{without neglecting less common ones} (\eg holidays such as \exinline{Christmas}).
Further, we want the process to be \textbf{cost-effective}, not wasting annotation effort on demonstrations that are redundant with already covered patterns.
To achieve these goals, we design \sysname to respond to every user interaction by offering assistance in three areas:
\begin{itemize}[leftmargin=1.2em,labelwidth=*,align=left]
\item \textbf{Help the user discover previously unexplored patterns}.
In each iteration, \sysname{} uses \emph{existing} demonstrations and users' past interactions to cluster the remaining unlabeled data into task-specific slices. Such slices map the input space for users to explore.

\item\textbf{Help the user prioritize the most informative examples}.
\sysname uses the current in-context function to estimate the difficulty of slices \emph{and} examples, prioritizing unexplored slices or slices and examples where the current function is not yet performing well.
We call this variant of active learning \emph{slice-based} sampling.

\item \textbf{Minimize annotation cost}.
Rather than providing a gold output label from scratch for each example, the user is presented with the best guess output from the \emph{current} in-context function (updated at every step), which they either accept when correct or \emph{edit} the incorrect parts.
\end{itemize}

We wrap these functionalities with a lightweight interface, where at each round, users are presented with a batch of unlabeled examples to be (potentially) added to the set of demonstrations. Thus, at each round, users get a ``picture'' of their current in-context function, and interact with it for improvement. We now detail each one of these components.

\begin{figure*}[t]
\centering
\includegraphics[trim={0cm 18cm 16cm 0cm}, clip,width=0.9\linewidth]{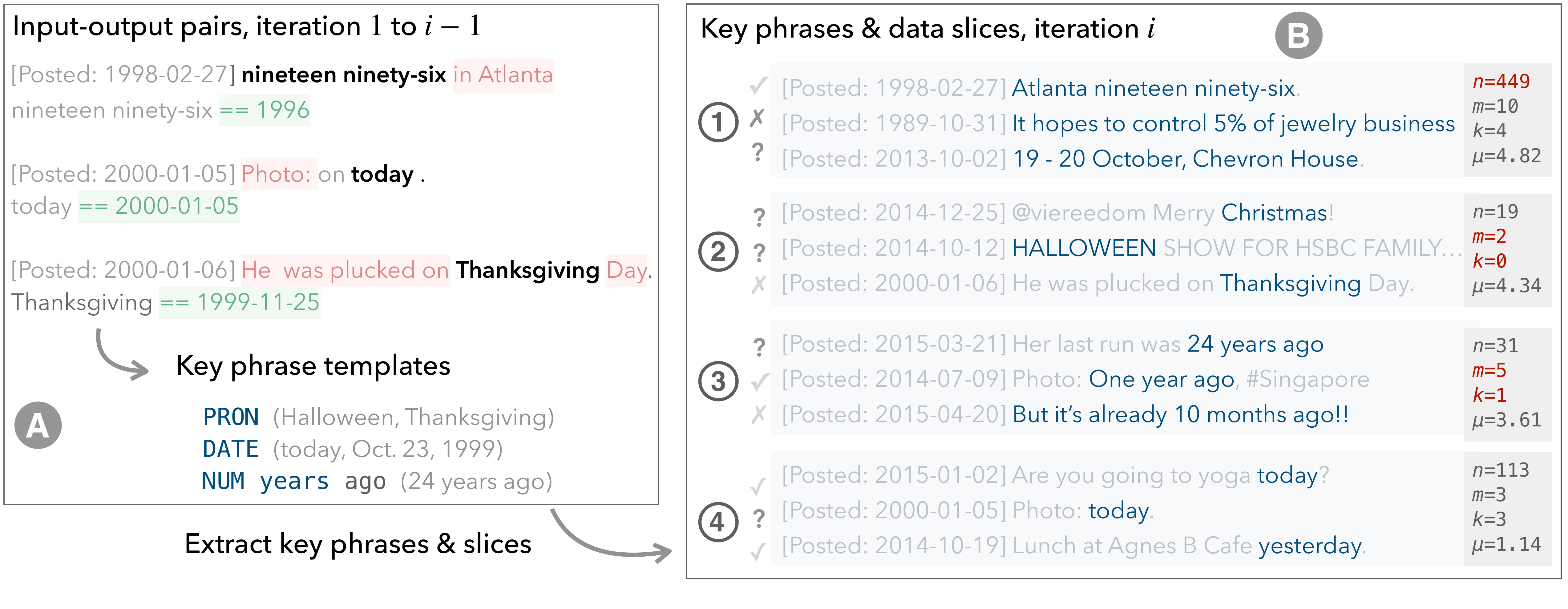}
\vspace{-5pt}
\caption{
An overview of \sysname's slice-based sampling. 
We use the data status from $1$ to $i-1$-th iteration to perform sampling for the $i$-th iteration.
As shown in (A), we use the already sampled input-output pairs to extract \emph{templates} for \emph{task-specific} key phrases.
We use these templates to extract key phrases for each unlabeled input, which are the blue highlights in (B).
For example, \texttt{PRON} helps extract \exinline{Christmas} from \exinline{@virreedom Merry Christmas!}. 
We run Agglomerative Clustering on the sentence embedding of these key phrases to find task-specific data slices, which contain both not-yet labeled examples (marked with ``?'') as well as those that have been sampled (``\cmark'' for correctly predicted in previous iterations and ``\xmark'' for incorrect predictions).
We rank these slices by an award function $\mu$ based on slice size, estimated model performance, and sample frequency, and draw samples from the top clusters.
}
%\vspace{-10pt}
\label{fig:slice}
\end{figure*}

\subsection{Interactive Interface} 
\label{subsec:interface}

We present \sysname as an interactive interface, shown in Figure~\ref{fig:ui}.
The interface contains a task description (A$_1$) and existing in-context examples as demonstrations, presented as input-output pairs (A$_2$).
These pairs are color-encoded based on the text editing distance, with the spans deleted from the input in red, and the spans added in green.
Both the description and demonstrations are editable, and are automatically translated into an LLM prompt (Figure~\ref{fig:ui}B) with the task description, demonstrations in the format \texttt{>> [example input] => [example output]}, and a candidate input for the LLM\footnote{All of our studies and experiments are run on GPT-3~\cite{brown2020language}, \url{https://beta.openai.com/}} to transform into an output. %\texttt{>> [candidate input] =>}, so the LLM can provide the potential output.

Below the existing examples, \sysname proposes a batch of $5$ \emph{candidate} inputs sampled from the unlabeled dataset, with outputs computed with the current version of the in-context function (A$_3$), using the prompt in Figure \ref{fig:ui}B. 
Users then verify the candidates and provide feedback (A$_3$), editing outputs to fix mistakes when needed (\eg changing from \exinline{Thanksgiving == 2000-11-25} to \exinline{Thanksgiving == \uline{1999}-11-25}, A$_4$), and adding or removing examples to the few-shot examples for in-context learning (A$_5$).
In addition to saving annotation time, LLM-generated outputs help users assess the quality of the current version of the in-context function.
For example, if all LLM outputs are correct for a few consecutive batches, it is likely that the existing few-shot examples cover the patterns in the unlabeled data, and thus labeling can stop.

\begin{figure*}
\centering
\includegraphics[trim={0cm 15.5cm 18cm 0cm}, clip,width=0.95\linewidth]{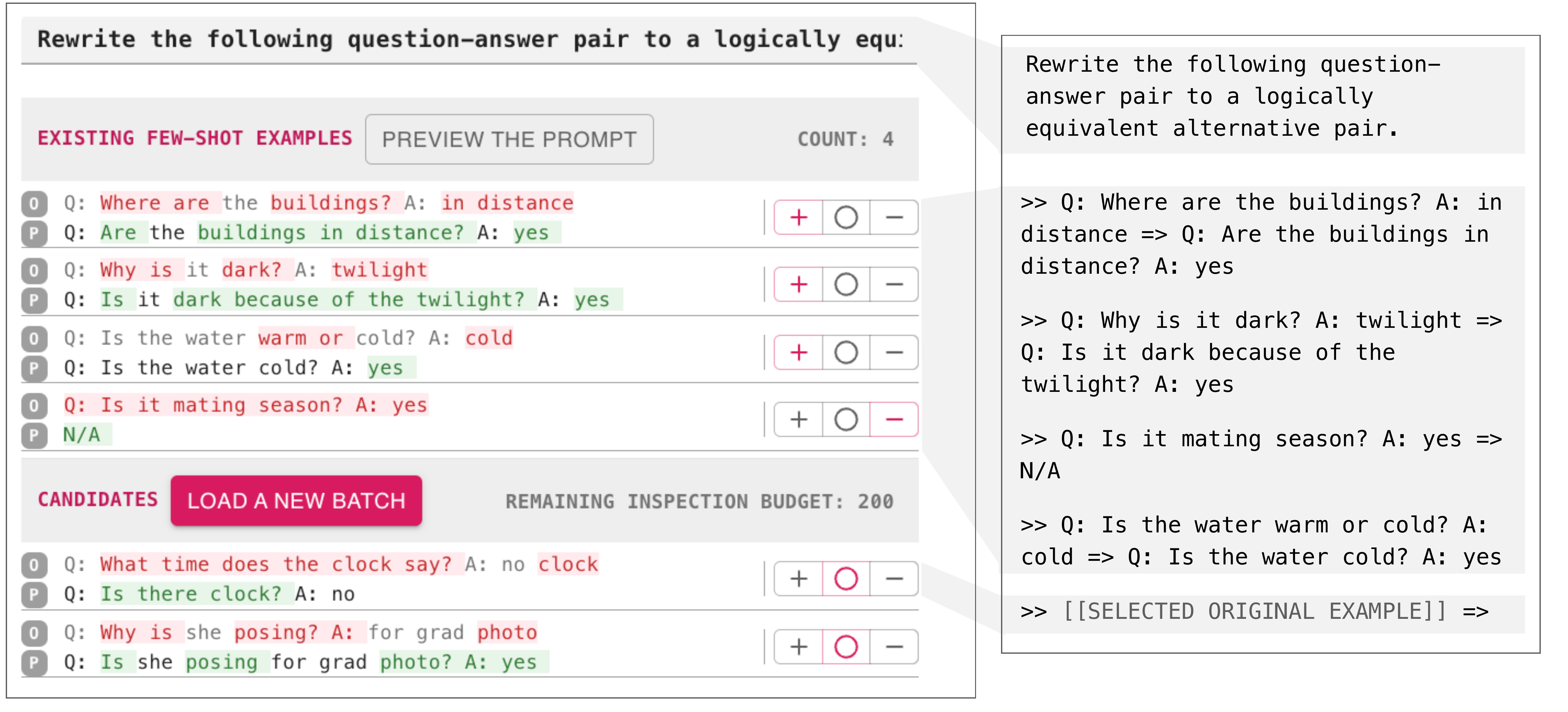}
\vspace{-10pt}
\caption{
The \sysname interface on the question-answer pair implication task.
}
%\vspace{-5pt}
\label{fig:ui_qa}
\end{figure*}

The interface is task-agnostic and can be used whenever users want to learn one-on-one text mapping between text inputs and outputs.
This format is flexible, encompassing both classification tasks (where the output is just the class name) and generation tasks like summarization, though the color encoding would be most effective for text transformation tasks where the edits from inputs to outputs are worth highlighting. For example, Figure~\ref{fig:ui_qa} shows how the same interface is used for another question-answer pair rewriting task.
%This format is flexible, encompassing extraction tasks (\eg Fig \ref{fig:ui}), transformation tasks (\eg Fig \ref{fig:ui_qa}, where the task is rewriting question-answer pairs), and classification tasks (where the output is just the class name).
% For example, Figure~\ref{fig:ui_qa} shows the interface for the question-answer pair rewriting task that we used for our simulation and user study.
\sysname{} can be easily invoked in a Jupyter Notebook, and therefore can support users' natural workflows.

\begin{figure*}[t]
\centering
\includegraphics[trim={0cm 26cm 23cm 0cm}, clip,width=0.8\linewidth]{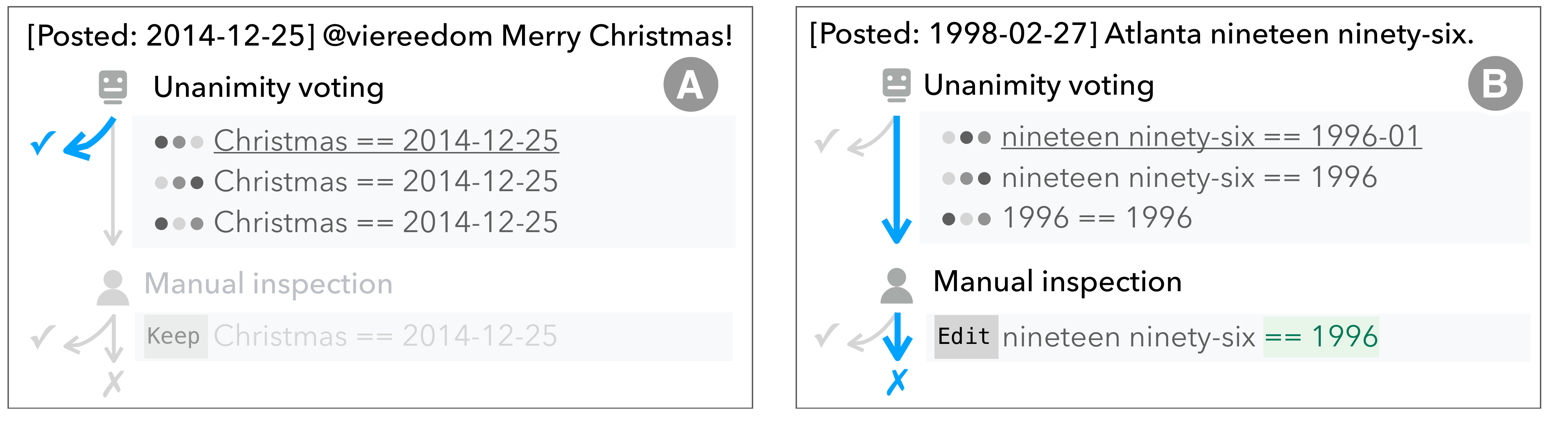}
\vspace{-10pt}
\caption{
An illustration of \sysname's two-step correctness estimation. 
When the in-context function demonstrates reasonable quality in the last two iterations, we first employ \emph{unanimity voting}, \ie we use three different orderings of in-context examples (noted with the three dots with different grey shades) to get three outputs, and say the function is correct if all the outputs are the same, without showing the input to the human (A). When the outputs are different, we return the one with the highest output probability for \emph{user inspection} (\uline{underlined}), in which manual editing would imply that the function is wrong (B).\looseness=-1
}
%\vspace{-5pt}
\label{fig:estimate_acc}
\end{figure*}

\subsection{Slice-based Sampling}
\label{subsec:sampling}

\subsubsection{Identifying patterns with key phrase clustering}
\mbox{}\\
To help users explore both \emph{common} and \emph{less common} patterns, we need a way to partition the unlabeled input examples.
While there are a variety of task-agnostic distance metrics that could be used for clustering (\eg cosine similarity of sentence embeddings~\cite{reimers-2019-sentence-bert}), our preliminary exploration indicated that these are typically too coarse when applied to \emph{specific} tasks.
For example, using the embeddings from \citet{reimers-2019-sentence-bert}, \exinline{Took a photo today.} is closer to \exinline{Saw a photo on Flickr.} (similarity $= 0.56$) than to \exinline{Are you going to yoga class today?} (similarity $=0.30$).
While this may make sense in the abstract, it does not correspond to how we would want to slice examples for the temporal extraction task in Figure \ref{fig:teaser}, where date references \exinline{today} are more important than subject matter (\exinline{photos} vs \exinline{yoga class}).
Thus, we propose a \emph{task-specific} clustering method based on key phrases as explained below.

\textbf{Detecting key phrases in demonstrations.} 
While key phrase extraction in general may require domain knowledge~\cite{2019-errudite, ratner2017snorkel, cabrera2022did}, for text transformation we can leverage the signal present in the relationships between input and output, \ie in which parts of the input are perturbed or retained. For example, \exinline{today} is retained in the output of both \exinline{Took a photo today.} and \exinline{Are you going to yoga class today?} (among many other samples), and thus it is probably a key phrase.
Formally, given a labeled, positive example, \ie a pair of original and perturbed sentences $f(x) => y$, we extract as key phrases either the unmodified parts of $x$ when most of $x$ is changed (Levenshtein edit distance $d(x, y) \geq 0.5$, as is the case with the ``today'' examples above), or the modified parts when most of $x$ remain unchanged.

\textbf{Applying key phrases to unlabeled inputs.}
Applying key phrases naively with an exact match would yield low coverage in the unlabeled data (especially for larger phrases).
To get more coverage, at each iteration, we generalize key phrases extracted from labeled demonstrations into \emph{templates} with combinations of tokens, lemmas, and part-of-speech tags \cite{wu2020tempura, wu-etal-2021-polyjuice}, \eg \exinline{today} is expanded into \texttt{today}, \texttt{NOUN}, and \texttt{DATE}.
We then select representative templates with a greedy weighted set coverage algorithm based on their specificity and the number of inputs they cover~\cite{vazirani2013approximation}.
Example templates at various abstraction levels are shown in Figure \ref{fig:slice}A.

\textbf{Key phrase clustering.} We define the distance between two inputs as the minimum cosine distance between the sentence embeddings~\cite{reimers-2019-sentence-bert} \emph{of their key phrases}, and use agglomerative clustering~\cite{lukasova1979hierarchical} to recursively merge pairs of clusters in the unlabeled data.
We set the number of clusters to $20$ (chosen empirically in Section~\ref{sec:simulation}), and aggregate all clusters with $<10$ examples into a single ``outlier'' cluster (Figure \ref{fig:slice}B$_1$).
Note that we recompute clusters in every iteration, and thus the outlier cluster tends to shrink as the user interacts with the system. Figure~\ref{fig:slice}B contains various examples of discovered clusters.

\reviews{R1: Slice-based sampling. Won’t the initial set of demonstrations (from which the templates are extracted) greatly impact the suggestions provided through slice-based sampling? Did this come up at all in the user study?}
\fixed{Note that as a result of the weighted coverage selection, the templates --- and thereby the extracted key phrases --- are dynamically changing, and will eventually become more dominant in the sampling procedure: 
when the few-shot set contains only a few (e.g., 3) seeding examples, the templates might be biased or even non-existent, most examples will just use the full sentences as key phrases, making it similar to vanilla clustering on full examples.
However, as we add more examples, the templates will be more balanced and eventually stabilize, at which point the clustering can rely more on the extracted key phrases.
}

\subsubsection{Selecting slices for exploration}
\mbox{}\\
We want to explore the identified slices in an efficient way, avoiding slices already ``solved,'' and making the user discovers any unexplored patterns.
We take inspiration from the UCB algorithm \cite{auer2002using}, and use an upper bound estimate of the error of our function in each slice as part of the ``reward'' for sampling from that slice.
Formally, suppose slice $c$ has $n$ examples, $m$ of which are labeled in previous iterations (see the next section for ``labeling'' details).
%--- \ie we have its groundtruth label, either because it predicted correctly in a previous iteration (see the estimation heuristic below), or because the human inspected and corrected the erroneous output.
Further, suppose that out of the $m$ previously labeled examples, the current function is correct on $k$.\footnote{If an example is in the in-context set, we perform cross-validation and predict its output using the remaining examples.} The reward of drawing from slice $c$ at iteration $i$ is then given by:
\begin{equation*}
\mu_{i,c} = \underbrace{(1-\frac{k}{m})  \vphantom{\sqrt{\frac{\ln{i}}{m}}}}_{\text{Error Rate}}
    \cdot \underbrace{\ln n  \vphantom{\sqrt{\frac{\ln{i}}{m}}}}_{\text{Size}} + \underbrace{\sqrt{\frac{\ln{\fixed{i}}}{m}}}_{\text{Sample Rarity}}
\end{equation*}
\reviews{R2: The ‘Sample Rarity’ in the reward function is not clear. The definition of ‘lnt’ in ‘Sample Rarity’ is not given under data slice ranking. [author: we mis-spelled ln i as ln t in the submission and has corrected it]}
In other words, we prioritize \emph{large slices} ($\ln n$),  \emph{low performance } ($1- k / m$), and slices that have not been sampled many times ($\sqrt{\ln{i}/m}$\fixed{, which would give higher weights to clusters with smaller $m$ as the iteration $i$ progresses}).
Thus, we avoid wasting annotation effort on slices that are already ``solved'', but keep drawing from slices we can't yet deal with and slices we have not yet explored.

Figure~\ref{fig:slice}B shows four data slices in temporal extraction ranked by reward $\mu$.
\textcircle{1} is the ``outlier'' cluster, where patterns are not yet apparent.
This slice still gets prioritized due to its large size ($n=449$), even though it has been sampled $m=10$, which encourages either higher accuracy or further slicing in follow-up iterations.
\textcircle{2} is a slice with holiday-based key phrases. Though the slice is small ($n=19$), the LLM failed whenever it was previously sampled ($k/m=0$), and thus it currently represents a hard pattern.
\textcircle{3} is a slice with past date references, while \textcircle{4} is a slice with the common temporal pattern represented by the words \exinline{today}, \exinline{yesterday}, and \exinline{tomorrow}.
This last slice has low priority despite being common, since the LLM had perfect accuracy whenever a sample from it was drawn.
To maximize diversity (similar to batched active learning~\cite{citovsky2021batch, sener2017active, geifman2017deep}), we rank the slices by reward and select one example from each until the batch is filled (in our case, batch size $=5$).%select one example for per slice from the top 50\% slices to return the batched examples to the interface, up to the batch size $b=5$.
%Below, we explain how we (1) identify data slices, and (2) estimate LLM performance on unlabeled inputs (\ie for computing $k$).

\subsubsection{Saving user effort with implicit labels and pseudo-labeling}
\mbox{}\\
As mentioned above, our per-slice performance estimation requires \emph{labeled examples}.
Unfortunately, we only have firm labels on user-added in-context examples, which may be quite small, especially if users only add a portion of the sampled data.
As a result, in-context examples offer limited power for estimation. 
Although we can modify the interface to collect additional user labels on output correctness, it requires additional interaction that can be cumbersome.
To save user effort, we use \emph{implicit labeling}, \ie we label the LLM output of an example in a batch as correct if the user does not make any changes to the output, even if they do not add it to the in-context demonstration set.
Of course, users might ignore model errors if they are frustrated or distracted, but we verified in pilot experiments that users almost always make corrections in the presence of model mistakes ($\sim$87\% of the time, and the selection method is robust to this small amount of noise).
In comparison to explicit labeling, this method requires the bare minimum user interaction, and is easier to integrate into iteration workflows.

Still, implicit labeling requires users to actually \emph{see and interact} with a sample.
However, after a certain point in the process, the LLM is correct often enough that many interactions would simply be ``accepted'' (no changes) by the user.
While important for estimating slice accuracy, too much of such interaction might also lead users to overestimate the in-context function quality, and stop the process before they explore the remaining slices.
Thus, after we reach a threshold of quality (LLM is correct on 70\% of examples in two consecutive rounds), we start leveraging pseudo-labeling with \emph{unanimity voting}, a method inspired by the  unanimity principle~\cite{khani2016unanimous} and Query-by-Committee~\cite{melville2004diverse}.
Following \citet{lu2021fantastically}'s observation that the order of in-context demonstrations can drastically change LLM performance, we form three different prompts by randomly reordering the examples.
When the outputs of the prompts agree (\ie are unanimous), we use that output as a \emph{pseudo-label}, used both for estimating slice accuracy and as a filtering method (\ie these examples are not shown to the user).
% The method also works as a filtering method; 
% If unanimity voting says the model is correct on an input, it is not shown to the human, so as to save their inspection budget.
Figure~\ref{fig:estimate_acc} illustrates this process, where \exinline{@viereedom Merry Christmas} (A) is pseudo-labeled due to unanimity, and \exinline{Atlanta nineteen ninety-six} (B) yields different predictions, and thus is shown to the user for manual inspection. \looseness=-1

\newcommand{\random}{\emph{Random}\xspace}
\newcommand{\scattershot}{\emph{ScatterShot}\xspace}
\newcommand{\temporal}{\emph{Temporal}\xspace}
\newcommand{\qa}{\emph{QA-pair}\xspace}

\renewcommand{\arraystretch}{0.9}
\begin{table*}[t]
%\small
%\fontsize{7.5}{8}\selectfont
\centering
\caption{
Quantitative results comparing \sysname with the random baseline on \temporal and \qa, averaged over 10 random seeds.
\sysname outperformed the baseline on all metrics.
The significant improvements, measured by student's t-test are marked with *: $p < 0.05$, and **: $p< 0.01$.
}
\label{table:sim_result}
\vspace{-10pt}
\begin{subtable}[ht]{0.6\textwidth}
\setlength{\tabcolsep}{3.5pt}
\begin{tabular}{@{}r l l l l l l @{}}
\toprule

\multirow{2}{*}{Conditions}
& \multicolumn{3}{c}{\textbf{Extraction}} & \multicolumn{3}{c}{\textbf{Normalization}} \\
\cmidrule(lr){2-4}
\cmidrule(lr){5-7}
& F1 & Precision & Recall & F1 & Precision & Recall \\
\midrule\midrule
Random &  73.2 $\pm$ 4.0 &  74.0 $\pm$ 3.8 &  72.9 $\pm$ 4.1 &  66.8 $\pm$ 3.2 &  67.3 $\pm$ 3.3 &  67.0 $\pm$ 3.1 \\
\sysname & \textbf{75.0 $\pm$ 2.9} & \textbf{75.6$\pm$ 2.8} & \textbf{74.7 $\pm$ 2.9} & \textbf{70.9 $\pm$ 3.4}** & \textbf{71.3 $\pm$ 3.5}* & \textbf{71.2 $\pm$ 3.2}** \\
\bottomrule
\end{tabular}
\caption{\temporal}
\label{table:sim_temporal}
\end{subtable}
\hskip 35pt {\color{white}\vrule} \hskip 10pt
\begin{subtable}[ht]{ 0.3\textwidth}
\renewcommand{\arraystretch}{1.0}
\setlength{\tabcolsep}{3.5pt}
\begin{tabular}{@{}r l l @{}}
\toprule

\multirow{1}{*}{Conditions}
& \multirow{1}{*}{ROUGE-L} & \multirow{1}{*}{BLEU-4}
%\hskip 13pt {\color{white}\vrule}
\\
\midrule\midrule
Rule-based &  78.4  &  66.7 \\
Random &  74.3 $\pm$ 3.9 &  65.4 $\pm$ 3.5 \\
\sysname & \textbf{80.0 $\pm$ 3.5}* &  \textbf{69.1 $\pm$ 3.1}*  \\
\bottomrule
\end{tabular}
\label{table:sim_qa}
\caption{\qa}
\end{subtable}
%\vspace{-10pt}
\end{table*}

\begin{table*}[t]
%\small
\centering
%\small
\renewcommand{\arraystretch}{1.1}
%\setlength{\tabcolsep}{pt}
%\fontsize{7.5}{8}\selectfont
\caption{
Example outputs from transformation functions built in \sysname and \random condition, and from a rule-based system \cite{ribeiro2019red}.
\sysname functions tend to have better coverage, fluency, and correctness.
}
\vspace{-10pt}
\begin{tabular}{@{} r | l | l @{}}
\toprule
\multicolumn{3}{l}{\textbf{Coverage}: Transforms more forms of inputs.}\\
\midrule
%\multirow{4}{*}{\rotatebox[origin=c]{90}{Coverage}}
Input 
    & Q: Are there more girls or boys? A: equal 
    & Q: How many hairs does the sheep in front have? A: infinite\\
Rule-based 
    & \xmarkcolor \emph{(No generation)}
    & \xmarkcolor \emph{(No generation)} \\
Random 
    & \cmarkcolor Q: Are the girls and boys equal in number? A: yes
    & \xmarkcolor N/A  \\
\sysname 
    & \xmarkcolor Q: Are the girls and boys equal? A: yes
    & \cmarkcolor Q: Does the sheep in front have infinite hairs? A: yes \\
\midrule\midrule

\multicolumn{3}{l}{\textbf{Fluency}: Generates outputs that sound natural.}\\
\midrule
%\multirow{4}{*}{\rotatebox[origin=c]{90}{Fluency}}
Input 
    & Q: What make is the phone? A: vtech
    & Q: What does the woman have on her face? A: headband \\
Rule-based 
    & \xmarkcolor Q: Make is the phone vtech? A: yes
    & \xmarkcolor Q: Does the woman have on her face headband? A: yes \\
Random 
    & \cmarkcolor Q: Is the phone a vtech? A: yes
    & \cmarkcolor Q: Does the woman have a headband on her face?  A: yes \\
\sysname 
    & \cmarkcolor Q: Is the phone a vtech? A: yes
    & \cmarkcolor  Q: Does the woman have a headband on her face? A: yes \\
\midrule\midrule

\multicolumn{3}{l}{\textbf{Correctness}: Produces desired outputs (the new question-pair are logically equivalent to the original pair).}\\
\midrule
%\multirow{4}{*}{\rotatebox[origin=c]{90}{Correctness}}
%& 
Input 
    & Q: What monument are they next to? A: unknown
    & Q: What type of motorcycle is in the picture? A: mountain\\
Rule-based 
    & \xmarkcolor Q: Are they next to unknown? A: yes
    & \xmarkcolor Q: Is the mountain in the picture ? A: yes \\
Random 
    & \xmarkcolor Q: Is the monument unknown? A: yes
    & \xmarkcolor Q: Is the mountain type of motorcycle in the picture ? A: yes \\
\sysname 
    & \cmarkcolor Q: Are they next to an unknown monument? A: yes
    & \cmarkcolor Q: Is the motorcycle in the picture a mountain bike? A: yes\\
\bottomrule
\end{tabular}
\centering
% \vspace{-5pt}

%\vspace{-10pt}
\label{tab:qa_qualitative}
\end{table*}
\section{Simulation Experiment: \sysname Sampling vs. Random Sampling}
\label{sec:simulation}

In this section, we measure the effectiveness of slice-based sampling, when compared to random sampling on two text transformation tasks.
We use datasets for which we have labels on both tasks, so that we can simulate the labeling process with an oracle at scale, and evaluate the learned function on a held-out portion of each dataset.\looseness=-1

% Is slice-based sampling helpful in revealing informative data points?
% We measure its effectiveness through simulation experiments on two text transformation tasks.
% We use fully labeled datasets in both tasks in order to simulate a labeling process with an oracle that always provides groundtruths, and quantify the learned in-context function quality by evaluating them on holdout sets.

\subsection{Tasks and Datasets}
\label{subsec:sim_task}

\noindent\paragraphBold{Temporal expression extraction and normalization}
The \textbf{\temporal} task involves data wrangling~\cite{verbruggen2021semantic}, where the goal is extracting phrases with temporal expressions from sentences or documents, and normalizing them into a standard format~\cite{chang2012sutime}.
As shown in Figure~\ref{fig:teaser}, these can include absolute or relative dates, and can have different granularity (\eg exact date vs. year only).

\textbf{Data.} We take the data from \cite{almasian2021bert}, containing temporal expression datasets, including TimeBank~\cite{pustejovsky2006timebank} (news articles) and TweeTime~\cite{tabassum2016tweetime} (tweets).
We process each dataset into sentences, discarding any date annotations that could not be normalized to the format YYYY-MM-DD (for consistency), and keeping sentences involving absolute dates, dates relative to the document publication date, or no time expressions at all (as the pool for negative examples).
This resulted in 491 examples with YYYY-MM-DD outputs, and $369$ negative examples with the output N/A.
We sample $100$ examples randomly from this dataset as a \textit{test set}, and use the remaining examples as our unlabeled pool in the experiment.

% Since some of the normalizations are arguably ambiguous and hard to control, we process the dataset such that (1) we only kept data points that could be normalized to the format of YYYY-MM-DD, and (2) we split long documents into single sentences, and only kept sentences that involve absolute dates, or describe dates relative to the document publication date, or no time expression at all (as the pool for negative examples).
% We collected 491 positive examples and 369 negative ones this way.
% \footnote{The processed data for both \temporal and \qa can be found in \url{\repourl}.}
% We randomly sample 100 examples to form the test set, and use the remaining examples as a pool for selecting in-context examples.

\textbf{Evaluation.} 
Following \citet{chang2012sutime}, we report F1, recall, and accuracy both for the temporal expression extraction and normalization separately.
%We run both selection strategies three times to account for variance.

% We follow \citet{chang2012sutime}'s practice in evaluating the result. Specifically, we report the F1, recall, and accuracy for both the time expression extraction and the normalization.
% We repeat the procedure three times for more stable observation.

\paragraphBold{Question-Answer Pair Implication}
For the \textbf{\qa} task, we use \sysname to replicate transformation functions from prior work.
Given a question-answer (QA) pair, \citet{ribeiro2019red} wrote a rule-based system (over $1,000$ lines of code\footnote{\url{https://github.com/marcotcr/qa_consistency/}}) to generate a new QA pair that is \emph{implied} by the original pair, to check whether question answering systems are consistent in their reasoning.
We replicate their \emph{logical equivalence} transformation, where the original QA is rewritten to a logically equivalent form, e.g. \exinline{Q: What room is this? A: bathroom} is transformed to \exinline{Q: Is this a bathroom? A: yes}.
Despite the heavy engineering, the rule-based system is not able to cover many inputs, and often produces text that does not look fluent or natural. We thus apply in-context learning to this task, and use \sysname{} to select the examples.

% In the \textbf{\qa} task,  we also explore using \sysname to replicate transformation functions in prior work for evaluating NLP model behavior consistency.
% In particular, we try to replicate the \emph{logical equivalence} transformation created by \citet{ribeiro2019red}, which rewrites a question-answer (QA) pair to its logically equivalent form: knowing the input QA pair is true would automatically implies that the output QA pair is also true, and vice versa.
% For example, if the input is \exinline{Q: What room is this? A: bathroom}, the output should be \exinline{Q: Is this a bathroom? A: yes}.
% These examples can help evaluate question answering models robustness, \ie a robust model should answer both the original and the perturbed question correctly (see more examples in Figure~\ref{fig:ui_qa}).
% %, and a sample clustering step in Appendix~\ref{appendix:qa_cluster}

% While the test objective seems intuitive, engineering such rewrites is often painful. For example, \citet{ribeiro2019red} had to create four subtly different templates that perturb different types of inputs based on their parsing tree structures. 
% Even with these, the templates were still too strict to generate natural texts for many diverse intended inputs.
% The task difficulty and the input diversity make it our ideal testbed.

\textbf{Data.}
We download the input sentences and rule-based implications from \citet{ribeiro2019red}, and manually inspect and label $1,000$ randomly sampled QA pairs ($351$ rule-based implications were noisy and had to be relabeled).
We randomly sample $100$ pairs as a test set, and use the remaining pairs as our unlabeled pool in the experiment.
% Since the template outputs are generally noisy and could not be directly used as oracle as in the temporal expression case, two authors manually inspected and fixed the outputs of 1,000 randomly sampled QA-pairs.
% In total, we re-annotated 351 examples.
% Similar to \temporal We randomly sample 100 examples to form the test set, and use the remaining examples as a pool for selecting in-context examples.
% To create the data space, we downloaded \citet{ribeiro2019red}'s pairs of input sentences and template-transformed outputs\footnote{\url{https://github.com/marcotcr/qa_consistency}}.
% Since the template outputs are generally noisy and could not be directly used as oracle as in the temporal expression case, two authors manually inspected and fixed the outputs of 1,000 randomly sampled QA-pairs.
% In total, we re-annotated 351 examples.
% Similar to \temporal We randomly sample 100 examples to form the test set, and use the remaining examples as a pool for selecting in-context examples.

\textbf{Evaluation.} 
We follow the standard in text generation and report the Rouge-L F scores~\cite{lin2004rouge}, as well as BLEU-4~\cite{lin2004rouge}.

\subsection{Procedure and Baseline}
\reviews{R1: Simulation experiment (Section 3). I thought this section could use a little more motivation on why the authors are comparing to random sampling. Is this a common practice reflected in prior works?}
We compare \scattershot's slice-based sampling with a \random sampling baseline, \fixed{which is the most common sampling method used especially in complex tasks, e.g., in text translation~\cite{agrawal2022context}.}
% Note that this is a stronger baseline than asking humans to manually construct a training set, as it assumes a reasonable access to less obvious cases that humans might have missed, and it also ensures that all the labels are correct.
%\hua{would it be helpful to slightly describe the random baseline's process? -- like random sampling from all data instances and filter out the correct one?}
We use GPT-3 as our underlying LLM, with greedy decoding (temperature=$0$) in both conditions.
%Specifically, we set temperature=0 and top $P$=1, to minimize variance caused by LLM uncertainty for nucleus sampling~\cite{holtzman2019curious}.
In each simulation run, we start the process with three random samples (the same for both conditions) of input-output. At every iteration, we compare the ground truth label with the candidate label proposed by the current in-context function. When the labels differ, we add the pair \texttt{(input, oracle output)} to the in-context example set, simulating the case where the user corrects a transformation and adds it to the set;
Otherwise, the oracle user does not perform any action, simulating cases where the user ignores examples where the current in-context function is correct.
% For each round of simulation, we start with the same three random seeding samples for \scattershot and \random, and use the groundtruth labels as oracle annotators. 
% Every iteration, we compare the groundtruth with the candidate label proposed by the current in-context function.
% If these two labels differ, we add the pair of \texttt{(input, oracle output)} into the in-context example set, simulating the case where the user corrects the functions;
% Otherwise, the oracle does not perform any action, simulating cases where humans decide to do nothing on examples that the current in-context function got right.

\reviews{R2: The procedure of simulation studies is not detailed, which may make readers confused. The details of applying different sampling methods should be clarified. [author: we believe this part has already detailed the the procedure, and felt there's not much to be clarified.]}
\fixed{The process is repeated until one of the following stopping conditions is met: 
(1) the in-context example set contains more than 40 data points (exceeding the LLM maximum context size),
(2) The oracle user has been presented with 100 examples (i.e. annotation budget is met),
(3) the in-context function provided the correct outputs in five consecutive iterations, or
(4) the in-context function's estimated accuracy for all slices of data is $\geq 80\%$.

We run ten simulation rounds with different random seeds, and report the (averaged) final function performance. We further track the function improvement trajectory over iterations on three randomly selected simulation rounds, by evaluating the intermediate in-context functions after every five examples are added.}

\subsection{Results}

As Table~\ref{table:sim_result} shows, \sysname's slice-based sampling outperforms the baseline on both tasks.
In \temporal, \sysname improves the $F_1$ for date span extraction by around 2 points, and the normalization by 4 points.
In \qa, \sysname outperforms \random by 6 points on Rouge-L, and even outperforms the heavily engineered rule-based system \emph{used to label most of the evaluation data}, despite needing $40$ or fewer in-context examples.
Table~\ref{tab:qa_qualitative} shows qualitative examples, where \sysname outperforms both baselines in terms of coverage, fluency, and correctness.
These results point to \sysname's potential on saving human efforts in creating fine-grained functions, alleviating the need for handcrafting templates.\looseness=-1
% Of course, we do sometimes see \random performing better; 
% Empirically, the results are affected by the quality of the three seeding examples, and the ``luck'' of the random sampling~\cite{lowell2018practical}.
% This echos prior work on the limitation of active learning, \eg the labeled examples are not drawn \emph{i.i.d} from the underlying data distribution~\cite{settles.tr09}, and therefore can sometimes be imbalanced~\cite{pop2018deep} or less effective than random sampling~\cite{imberg2020optimal}.
% Still, the fact that \sysname sampling leads to better average function performance \emph{even when we have up to 40 examples} is encouraging.

\begin{figure*}
\centering
\includegraphics[trim={0cm 18cm 21.5cm 0cm}, clip,width=0.95\linewidth]{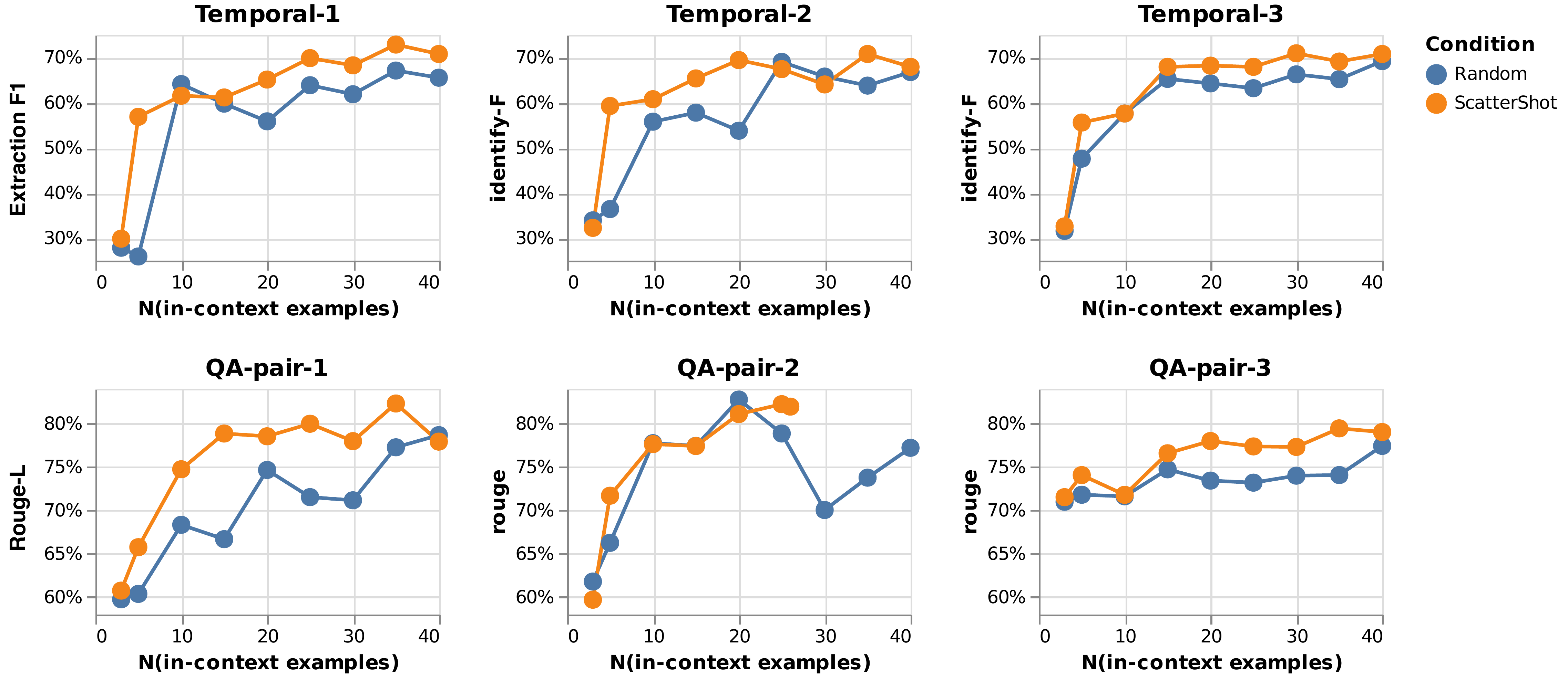}
\vspace{-5pt}
\caption{The in-context function performance trajectory, 
We evaluate the in-context function on the held-out test set every time we add five more examples to the in-context bucket until the stop condition is satisfied.
\sysname tends to frequently outperform \random, and tends to have less performance oscillation.
Note that the y-axis is different for \temporal and \qa.
}
\vspace{-5pt}
\label{fig:simulate_traj}
\end{figure*}

Figure~\ref{fig:simulate_traj} shows the trajectory of the in-context function quality as the simulated user adds more examples, for three randomly selected runs.
\sysname dominates the baseline at almost all points in all runs, with the biggest gaps occurring when the number of in-context examples is small.
We see particular gains at $n=5$, i.e. when the first two examples are added to the seed examples. 
Our hypothesis (based on qualitative observation) is that \sysname consistently selects examples that represent \emph{patterns} not contained in the seed examples, \eg negative examples (where the outcome is N/A) when all seed examples are positive.
While \sysname helps users explore most patterns in the unlabeled data as they reach higher $n$, early gains are especially useful in practice when users have low annotation budgets, \eg prior work notes users selecting as few as five or ten examples \cite{lu2021fantastically, park2022social}.

Finally, we observe that \sysname is less liable to variance in quality as more examples are added (e.g. in \qa-2, baseline performance degrades by almost 15 points between $n=20$ and $n=30$).
These results suggest that besides its interface and interactivity benefits, \sysname can improve in-context learning just by virtue of its sample selection function.
In order to evaluate the benefits to actual humans, we now turn to a user study.

% This is especially useful, as prior work has observed that in practice users may only provide as few as ten (or sometimes five) in-context examples~\cite{lu2021fantastically, park2022social}.
% Interestingly, the gain is especially noticeable at $n=5$, \ie when we just add two more examples to the identical seed examples. 
% We suspect this reflects \sysname's diversity selection.
% For example, in \temporal, \sysname would consistently present a negative example in the first iteration (\eg \exinline{[Posted: 1999-06-07] Reno said Bin Laden and Kopp have one thing in common. => N/A}) when the existing three seeds are all positive.
% We further plot the trajectory of the in-context function growth, as shown in Figure~\ref{fig:simulate_traj}.
% In general, \sysname is compatible or outperforms \random at most checkpoints, especially the earlier ones. 
% This is especially useful, as prior work has observed that in practice users may only provide as few as ten (or sometimes five) in-context examples~\cite{lu2021fantastically, park2022social}.
% Interestingly, the gain is especially noticeable at $n=5$, \ie when we just add two more examples to the identical seed examples. 
% We suspect this reflects \sysname's diversity selection.
% For example, in \temporal, \sysname would consistently present a negative example in the first iteration (\eg \exinline{[Posted: 1999-06-07] Reno said Bin Laden and Kopp have one thing in common. => N/A}) when the existing three seeds are all positive.

% In the next section, we look into \sysname's impact on human annotators in more detail.

\section{User Study}
\label{sec:user_study}

\sysname sampling is effective in simulation, but does it actually aid humans to articulate their desired functions? 
We conducted a within-subject user study to evaluate whether human users can sense \sysname's support in exploring the data space. 

\subsection{Study Design}
\label{subsec:study_design}

\newcommand{\smanual}{\emph{Manual}\xspace}
\newcommand{\spassive}{\emph{Random}\xspace}
\newcommand{\sactive}{\emph{ScatterShot}\xspace}
\newcommand{\cPtoA}{\emph{M-R-S}\xspace}
\newcommand{\cAtoP}{\emph{M-S-R}\xspace}

\newcommand{\cPtoAfull}{\smanual-\spassive-\sactive\xspace}
\newcommand{\cAtoPfull}{\smanual-\sactive-\spassive\xspace}

\newcommand{\sMtoP}{\emph{Manual$\rightarrow$Random}\xspace}
\newcommand{\sMtoA}{\emph{Manual$\rightarrow$ScatterShot}\xspace}
\newcommand{\sPtoA}{\emph{Random$\rightarrow$ScatterShot}\xspace}
\newcommand{\sAtoP}{\emph{ScatterShot$\rightarrow$Random}\xspace}

\paragraphBold{Task \& Participants}
We ran a user study on the \qa task using the same dataset as Section~\ref{subsec:sim_task}, with a split of 900 unlabeled inputs for participants to access, and 100 test examples for evaluating the in-context functions they built.
We recruited ten CS graduate student participants (4 females, 6 males) on our CSE department mailing list. Eight of them had previously used GPT-3 and two had heard about it, but none were familiar with the task or \sysname.
Each participant spent around 60 minutes in the study.\looseness=-1

\paragraphBold{Settings \& Conditions}
In order to isolate the effect of the different components in \sysname, we have two ablation settings in addition to our method:
(1) \textbf{\smanual}, where participants manually craft prompts without any help from \sysname, which is the de-facto \emph{status-quo} of practitioners creating their own in-context learning examples.
(2) \textbf{\spassive}, where participants use the \sysname interface with slice-based sampling disabled, \ie they review randomly selected examples. This condition still has the benefit of an interactive interface, and uses the intermediate in-context functions to suggest outputs and pseudo-label.
%This condition emphasizes on the access to lastest in-context functions and the \sysname interface.
%In this condition, users still receive annotation from LLMs, but not suggestions on feedback-worthy inputs.
(3) \textbf{\sactive}, where participants have access to \sysname, fully featured.\looseness=-1

Every participant interacted with every setting in sequence and in a cumulative manner, \ie the in-context demonstrations gathered in one setting carry over to the next, and we measured the \emph{additional} benefit of moving to the next setting.
We divided the participants into two groups, such that in one group the sequence is \smanual $\rightarrow$  \spassive $\rightarrow$ \sactive (\textbf{\cPtoA}), while in the other it is  \smanual $\rightarrow$  \sactive $\rightarrow$ \spassive (\textbf{\cAtoP}).    
\cPtoA represents a condition where participants are gradually exposed to more features, such that the step-wise gain maps directly to the benefit of the new feature, while \cAtoP serves as the counterbalanced condition that combats the learning effect and the natural impact of accumulating examples on function qualities.

\begin{figure*}
\centering
\includegraphics[trim={0cm 26cm 11cm 0cm}, clip,width=1\linewidth]{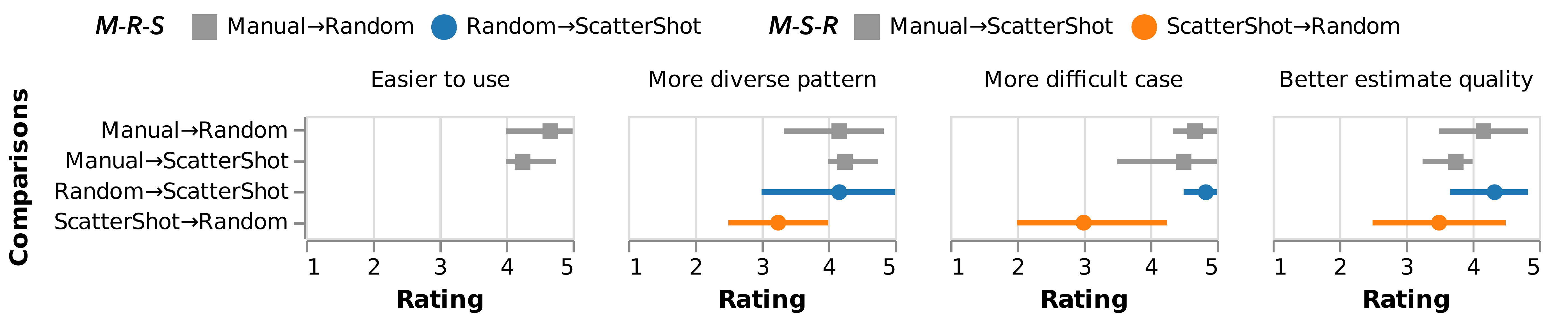}
\vspace{-15pt}
\caption{
Participants' subjective ratings on their perceived differences between different settings as they switch between them. 
We use the rectangle to represent when participants first move from \smanual (Step 1) to the \sysname interface (either \sactive or \spassive, Step 2), and circles to represent switches between \sysname interfaces, from one sampling method to the next (Step 2 to 3).
Participants strongly preferred the \sysname interaction to manual example annotation, and felt they found more diverse patterns and difficult cases in \sactive than \spassive (\sPtoA, blue).
In contrast, people in the reversed condition did not find \spassive more useful than \sactive (orange).
}
\vspace{-5pt}
\label{fig:likert}
\end{figure*}

\paragraphBold{Study Procedure.}
We designed our hour-long study to be self-contained in a Jupyter Notebook,\footnote{The full user study instructions, as well as the detailed exit survey, are in \url{\repourl}} and one of the authors was present in all studies to ensure that participants understood the task and to answer any questions.

Participants were first introduced to the basic concepts of LLM (GPT-3), in-context example construction, and the study task. Then, we randomly assigned the participants to one of the two conditions (\cPtoA or \cAtoP), and they completed the task by going through the three conditions in the assigned order. 
Participants were not instructed on the difference between  \sactive and \spassive, and were instead told that ``these two selection methods are randomly ordered, and one is not necessarily better than another.''

In each step (setting), participants were told to inspect the inputs and current function outputs (available in \sactive and \spassive), fix the erroneous outputs, and add demonstrations (input-output pairs) to the in-context example bucket if they believed the data would add additional value, \eg instances where the current context function fails, as well as diverse input or output patterns.
They were asked to iterate within the step until they were satisfied with the in-context function at hand, or accumulated 40 examples.
To prevent them from stopping too early, we also asked them to run at least three batches (\ie see 10-15 examples).\footnote{In \smanual, this meant looking at three random batches of unlabeled data in the Jupyter notebook.}
Afterwards, participants completed an exit survey and a semi-structured interview, where they rated their perceived experience in  each of the two consecutive steps.
These questions concerned their perceived input/output pattern diversity, the example difficulty, and their confidence in estimating in-context function quality.

\paragraphBold{Collected Data.}
We observed and analyzed three sets of data.
First, to quantify the change in \textbf{function quality}, we saved participants' in-context examples per step, and applied them to the held-out test set. 
Here, besides the absolute numbers as in Section~\ref{sec:simulation}, we calculated the \emph{difference} in performance between two consecutive steps to see if adding (or, in the case of \cAtoP, removing) \sysname features impacted the quality of examples participants submitted.
%This evaluation will be similar to the case study simulations, only with a more realistic baseline with actual humans rather than oracles.
Second, to assess participants' \textbf{self-perceived experience}, we used a standard five-point Likert Scale~\cite{likert1932technique} to collect their perceived step-wise differences.
Third, to track participants' \textbf{annotation trajectories}, we logged their clickstreams in all the steps.
This included both the number of examples they examined per step, the edits they made, and the number of examples they added.\looseness=-1

\subsection{Results}
\label{subsec:user_result}

\renewcommand{\arraystretch}{0.9}
\begin{table*}[t]
\caption{
The performances of participants' in-context functions after each step.
+/- represents the average performance change compared to the prior step, whereas the number in the parentheses are the absolute performances. \cPtoA participants were able to keep adding useful examples, whereas \cAtoP participants \emph{decreased} the function performance by 0.6 in Step three (\sAtoP), indicating that these efforts were wasted.
}
\vspace{-10pt}

%\small
%\fontsize{7.5}{8}\selectfont
\centering
\begin{subtable}[ht]{ 0.4\textwidth}
\begin{tabular}{@{}r l l l @{}}
\toprule

\textbf{Condition} &\textbf{Step 1} & \textbf{Step 2} & \textbf{Step 3} \\
%\hskip 13pt {\color{white}\vrule}

\midrule\midrule
\cPtoA &  / (59.3) & +17.4 (74.7)  &  \textbf{+3.2}  (77.8) \\
\cAtoP &  / (61.8)  &  \textbf{+18.1} (75.4) & -0.4 (74.9) \\
\bottomrule
\end{tabular}
\label{table:user_rouge}
\caption{ROUGE-L}

\end{subtable}
\hskip 30pt {\color{white}\vrule} \hskip 0pt
\begin{subtable}[ht]{ 0.4\textwidth}

\begin{tabular}{@{}r l l l @{}}
\toprule

\textbf{Condition} &\textbf{Step 1}  & $\rightarrow$ \textbf{Step 2} & $\rightarrow$ \textbf{Step 3} \\
%\hskip 13pt {\color{white}\vrule}
\midrule\midrule
\cPtoA &  / (63.9) & \textbf{+10.1} (74.0)  &  \textbf{+3.1}  (76.9) \\
\cAtoP &  / (65.3)  &  +8.9 (74.2) & -0.6 (73.6) \\
\bottomrule
\end{tabular}
\label{table:user_bleu}
\caption{BLEU-4}

\end{subtable}

\label{table:user_performance}
\end{table*}

% In analyzing subjective ratings, the logged clickstreams, as well as the final collected in-context examples from participants, we found:

\paragraphBold{The \sysname interface made it easier to iterate on in-context examples.}
As shown in Figure~\ref{fig:likert}, participants' found moving from \smanual (Step 1) to a \sysname interface (Step 2) beneficial, regardless of the sampling setting.
In particular, they found that the interface made it easier and more intuitive to construct the few-shot examples. (\textbf{Easier to use} in Figure~\ref{fig:likert}, 4.7 $\pm$ 0.7 for \sMtoP and 4.2 $\pm$ 0.4 for \sMtoA). 
Users liked the fact that \sysname offers sample inputs (rather than having to go through the dataset on their own), and the that the interface provides easy access to all the existing in-context examples, allowing for fast back-and-forth iteration. 
For example, one participant (P7) kept revisiting their examples, and removed some earlier examples that they thought were less useful as they became more familiar with the unlabeled input space.

As part of the interface, LLM-generated outputs helped participants craft examples more efficiently, \eg P6 comments that \quoteinline{it is less work to make edits than starting from scratch.}
Somewhat surprisingly, LLM-generated outputs also improved \emph{output diversity}, \ie users considered more diverse output patterns. 
For example, P10 commented that they were \quoteinline{pleasantly surprised by the LLM's clever output in several cases,} and that they would not have thought about transformations such as \exinline{Q: Is there more than 1 boy? A: no} $\rightarrow$ \exinline{Q: Is there no more than 1 boy? A: yes}, which they added to their set of in-context examples.
The observation is consistent with prior work showing AI-induced creativity gains~\cite{wang2022interpretable}.
We note that actual user behavior here differs from our simulation setup, where we assumed human users would \emph{only} add new examples when the LLM output was wrong.%, and demonstrates extra benefits from the human perspective.

\paragraphBold{Participants' perceptions matched \sysname's slice-based sampling design goals: more diverse and more challenging patterns.}
As shown in Figure~\ref{fig:likert}, participants in \cPtoA clearly noticed the improvement moving from \sPtoA (4.2 $\pm$ 1.2 for \textbf{more diverse patterns} and 4.8 $\pm$ 0.4 for \textbf{more difficult case}), whereas most users in \cAtoP did not report improvements from \sAtoP.
Qualitative results confirm this, \eg P7 in \cPtoA commented: \quoteinline{Step 2 (\spassive) provided me with some worthy examples, but much less than Step 3 (\sactive). I went through several rounds of pretty similar examples, thinking the function is behaving quite decently, and didn't realize the function needed more diverse and edge cases until I reached Step 3.}
P9 in \cPtoA was also happy that \sactive helped them explore beyond typical patterns. 
In contrast, P10 in \cAtoP reflected that their exploration seemed to have \quoteinline{quickly saturated in Step 3} (\spassive). 
% Out of pure ``luck'', most examples they received during Step 1 and 2 (\smanual and \spassive) were negative examples with an \exinline{N/A} output, and they thought they \quoteinline{get diverse patterns only at the end}.
%P8 in \sAtoP assumed Step 3 (\spassive) was our intended contribution and even felt sorry when they did not rate \spassive more useful.

Despite not being given details, seven participants discerned the goals behind \sysname's sampling method by interacting with it.
% This infers that \emph{\sysname selection is aligned with human intuition.} 
For example, P2 described it as \quoteinline{sample for additional variation based on the patterns in existing examples, and also sample for examples similar to previous error mistakes to track whether the function has been corrected.}
Two participants in \cAtoP noticed that \spassive presented fewer mistakes, but attributed it to the increasing number of in-context examples (P5: \quoteinline{It's getting more correct, but I would expect it given that I have annotated more examples}).
After we explained the selection methods at the end, some users noted that understanding the methods would have helped them better calibrate their estimates of the learned function quality over time.\looseness=-1
%expectations on the examples independent of their current in-context example size.

\paragraphBold{\sysname helped participants explore the input space more holistically, and build better in-context functions.}
The perceived data difficulty and diversity encouraged participants to iterate more on their in-context examples. 
When looking at the number of in-context examples added in each setting, participants added 40\% more examples in \sactive than \spassive when \sactive came after (\cPtoA), and 20\% \emph{fewer} examples in \spassive when \spassive came after (\cAtoP), \ie they stopped much earlier when \spassive came after \sactive.
% \sAtoP participants added 140\% more examples in \sPtoA than their own \sMtoP additions on average ($n=12.3 \pm 6.1$).
% In contrast, \sAtoP participants stopped much earlier, and only added \textbf{80\%} examples in \sAtoP than \sMtoA.
These additional examples are not only a result of more inspection effort  (on average, participants in \sactive reviewed 20\% more samples), but also that each batch in \sactive was more likely to contain a good in-context example --- participants added 81\% of the examples they inspected in \sactive, but only 48\% of the examples in \spassive.
% The examples they added in \spassive were also not all informative. 
% P8 described such tendency as avoiding \quoteinline{sunk costs}, saying that they were inclined to add an example \emph{even when the LLM already got it correct}, just because they have already spent the effort reading the example.
% Unfortunately, by doing so, they also wasted the computation budget running LLMs. 
% This points to the importance of allocating human inspection budgets towards more informative examples.

We report the \emph{quality} of the resulting in-context function on the held-out set in Table \ref{table:user_performance}, and note that \sPtoA consistently increases performance, while \sAtoP consistently \emph{decrease} performance despite adding more in-context examples, which is in line with our simulation results.
% The \sPtoA examples indeed appeared to be more effective. 
% As shown in Table~\ref{table:user_performance}, the additional examples in \sPtoA increased the in-context function Rouge-L by 3.1 $\pm$ 3.7, whereas those in \sAtoP \emph{decreased} the function performance by -0.6 $\pm$ 1.4, meaning that adding more random examples after \sactive could be an wasted effort. 

\paragraphBold{\sysname helped participants estimate function quality and ``debug'' their example set.}
As expected, participants estimated their in-context function quality based on the candidate examples they reviewed.
For example, P5 (\cAtoP) tracked the function convergence: \quoteinline{I made mental notes on the LLM errors and hypothesized what types of examples were missing. For example, I noticed the model was wrong on \texttt{N/A} questions at first, but later got it right.}
Participants in \cPtoA seemed slightly more satisfied with their estimation, with 4.2 $\pm$ 0.9 in \sMtoP and then further 4.3 $\pm$ 0.7 \sPtoA.
P7 commented that \quoteinline{Step 2 showed me the function is quite smart on patterns it has already seen and has high precision, and Step 3 showed me there are more patterns and it has low recall}.
P2 further reflected that \spassive's sampling \quoteinline{created a false impression of convergence, when the function still had various blind spots.}
The interactive process also helped participants debug their example sets, \eg P4 saw big performance drops (4/5 to 1/5 accuracy) on two consecutive batches, which led them to remove in-context examples that were hurting performance.
% They hypothesized that their examples added in the performance churning round were bad, and subsequently removed them.

Participants in \cAtoP gave slightly lower ratings on their estimates.
Qualitatively, the fact that \sactive prioritized potential mistakes seemed to discourage users, \eg P3 noted they were driven into \quoteinline{an endless blackhole of errors,} after which a round of repetitive patterns in \spassive was hard to make sense of.
% However, \cAtoP participants seemed slightly less impressed for two reasons: 
% (1) their very first interaction with the interface (\sactive) presented them with various LLM mistakes that made the participants think the LLM is unpredictable, 
% (2) they saw more repetitive patterns in \spassive that did not make it more helpful.
% In particular, directly jumping into \sactive seemed to create some frustration for people, as it drove participants intoyP3) --- this also means the debugging capability is somewhat mitigated in \sactive
Once again, this could have been mitigated by \emph{explaining} the sampling strategy to the users, and explicitly displaying the slice accuracy estimates \sysname keeps track of.
% Alternatively, 
% This points to potential revisions on the design of human-AI collaborative labeling, where the AI should balance the example informativeness with the sense of progress from the human perspective (discussed more in the next section).
\section{Discussion}

In this work, we design a human-LLM collaboration mechanism in \sysname to help humans craft diverse and informative in-context learning examples.
By iteratively identifying data slices, sampling from low-performance or unseen slices, and providing best-guess outputs for the sampled examples, \sysname not only helps the collection of informative in-context examples, but also supports users in exploring the input space and assessing the function quality.
At its core, \sysname is built on three concepts: data slicing and sampling, iterative human-model interaction, and collaborative human-model labeling. 
We now discuss challenges and potential future work for each of these.

\paragraphBold{Slice-based sampling can increase data space coverage.}
Our experiments showed that sampling from diverse and difficult data slices improves in-context function performance.
Importantly, these slices cannot be surfaced via clustering on task-agnostic embeddings; rather, task-specific features should be considered to group examples, while task-irrelevant noise should be minimized.
However, identifying these task-specific features remains a challenge. 
\reviews{Template-based extraction. As the authors note, this will only work on a few, select tasks. I think the paper would benefit from a more in-depth discussion on how to extend ScatterShot to other tasks, like translation. we believe we have covered this in the original submission.}
\fixed{While effective for our function examples (and many others), key-phrase and template extraction would not generalize to tasks where input and output have little syntactic overlap, \eg English-French translation, summarization, etc.
% In many tasks, the data space can even be sliced differently based on the selected feature~\cite{2019-errudite}.
Future work should look into incorporating more general slicing methods, \eg asking practitioners for slicing functions~\cite{chen2019slice, ratner2017snorkel, 2019-errudite}, automatically detecting blind spots~\cite{sagawa2019distributionally, eyuboglu2022domino}, etc.
}

In addition to data slicing, the sampling algorithm also plays a crucial role in narrowing down the actual slices to sample from.
We adapt the UCB algorithm to prioritize slice size, performance, and sample rarity, but there are other interesting dimensions that could be explored.
For example, if there are slices that cannot be learned after several rounds of sampling, UCB may be counterproductive and create a biased in-context example set that performs worse on \emph{other slices}, whereas a strategy that penalized or just ``gave up'' on those slices might produce a better overall function.
% For example, if some slices are so hard that it could not be grasped even after several rounds of sampling from them, then we might want to stop early on the corresponding slices in order to avoid over-biasing the in-context samples. 
% In those cases, we might want to penalize slices with minimum \emph{delta} accuracies across iterations.
Moreover, we might want to explore better methods for example ranking \emph{within a slice}.\looseness=-1

%intra-slice example ranking is also worth exploring, such that we do not perform sub-optimal exploration on a chosen slice.
% 
% There are ongoing efforts in creating datasets slice discovery\footnote{\eg \url{https://dcbench.readthedocs.io/en/latest/tasks.html\#slice-discovery}}, with a focus on images and structured data.
% An extension of such datasets to NLP would probably help the would greatly help track our process on the slicing space.

\paragraphBold{Interacting with the latest function is essential for in-context learning.}
In-context learning enables rapid function updates, which are not possible in other current interactions with models (\eg finetuning often takes long hours, and is often not suitable for interactivity).
Allowing users to interact with the most current version of what is being learned helps them track progress, and backtrack when they introduce cascading errors~\cite{jiang2022promptmaker}.
The setup in \sysname is a step in this direction, since users always interact with the latest version of the in-context functions.% and  and making estimates on function qualities on different data slices.

\reviews{R1: One feature I think the system is missing is a sense of an overview of the data the user has seen or sampled from the input space. The author’s touch upon this in the Discussion, where they discuss presenting quality metrics to the user to show them how their prompt is performing. How would you show this overview, or rather, visualize the user’s progress in improving the prompt?}
While participants \emph{were} making progress with \sysname (more than with baselines), some participants felt frustrated by inspecting mistake after mistake, fearing that they would never be able to produce a good enough function.
While this is by design (\sysname prioritizes potential errors), it might compromise annotators' estimates of the quality of their function, and their motivation for labeling more examples.
Thus, we notice the importance of presenting quality metrics to the user and clearly explaining the sampling function so that the right expectations are set.
For example, users may perform better mental calibration if they have access to hints like the number of slices that are considered ``solved'' \fixed{(e.g., as a progress bar that allows people to zoom into concrete examples grouped by the slice)}, cross-validation accuracy on in-context examples, etc.
Another alternative would be to let users exercise \emph{more control} over which slices are explored, \eg allowing them to ``drill down'' or ``give up'' on specific slices.
% This is arguably an expected side effect of the \sysname design, whose sampling and instability filtering reduces the chance of presenting correctly predicted examples. 
% However, this frustration compromises annotators' ability to estimate their function at hand, their expectations of LLMs, and their motivation to further iterate on their functions (``Am I making the function worse?'').
% The observation suggests that it may be necessary to achieve informative iterations without hampering human confidence. %by balancing the correct and incorrect examples presented.
% This requires augmenting \sysname with more sophisticated human-AI collaboration design and hints.
% For example, even if the interface does not present correct examples, it can display a number indicating that the new batch of \emph{errors} were found after \emph{a large number of correct examples}, so as to hint on the function correctness.
% Further, the interface can display the number by slice, so annotators can explicitly track the per-slice performance~\cite{2019-errudite}.
% In fact, this is quite related to our participant's suggestion on getting more rationales for each examples selected: we can reveal which slice an example was selected from, and whether it is because the slice as a whole is under performing, or the example is an edge case in a large slice.
% More studies on what information to disclose and how to display them would be promising.

\paragraphBold{Human-AI collaborative labeling for building better functions with respect to better quality \emph{and better task definition}.}
Essentially, \sysname enables human-LLM collaboration on data annotation.
In our work, we mostly focused on evaluating the quality benefit of such annotation, but we observed additional interesting gains in bringing people inspiration.
In Section~\ref{sec:user_study}, we notice that participants can take inspiration from the LLM not only on the input patterns, but also on potential output patterns even though our \qa task is relatively deterministic in its transformations.
Thus, we hypothesize that similar systems supporting human-LLM collaborative labeling could play an important role in helping users iteratively refine their task definition and function behavior during data collection.
Prior work has shown that annotation requesters refine their labeling instructions when they see noisy (and therefore unusable) crowdsourced labels on ambiguous examples. 
However, we have yet to examine how LLMs' suggestions (good or bad) might help users better specify their functions.%will shift their definition.
It would be interesting to systematically analyze and measure users' own distribution shift as the example set expands.
Recently, \citet{lee2022coauthor} proposes the ``retaining rate'' of LLM suggestions (in their case, suggested character names subsequently used in novels) as a metric of the usefulness of LLMs for ideation.
An analogue to our case would be measuring the appearance of new patterns \emph{data slices} when users use \sysname, compared to when they come up with their own patterns.
%We could similarly build metrics around the appearance of new information, \eg number new data slices after certain amount of additional labeling.

\section{Related Work}

\subsection{LLMs and In-context Learning}

Transformer-based large language models (LLMs)~\cite{vaswani2017attention} have recently led to large improvements in NLP.
Pre-trained on a large amount of unlabeled text data, these models encapsulate rich,  general-purpose features of language both syntactically and semantically. 
These features can help facilitate various downstream applications much more dynamically~\cite{liu2021pre} --- rather than having to train a new model for every custom task, users can just customize the model by feeding it natural language \textbf{prompts} at run time, like the holiday in the previous section. 
Such ability to recognize the desired task on-the-fly is called \emph{in-context learning}~\cite{brown2020language}. 

The flexible in-context learning intrigues various work to explore designing prompts that can effectively invoke the user desired functionalities~\cite{min2022rethinking, xie2021explanation, rubin2021learning, mishra2021cross}. 
To date, the most common patterns for prompting are either \emph{zero-shot} or \emph{few-shot} prompts.
Zero-shot prompts directly describe what ought to happen in a task. 
For example, we can enact the holiday date translator in Section~\ref{sec:intro} with a \emph{task description} prompt: \exinline{Identify the date for a national holiday in the month/date format.}
Studies on improving zero-shot prompts typically study the effect of task instructions~\cite{efrat2020turking}, induce LLM reasoning through task decomposition~\cite{wu2022ai, wei2022chain}, etc.
Zero-shot prompts do not use demonstrative examples and therefore tend to be less performative~\cite{brown2020language}, but writing just the natural language descriptions is lightweight enough that it creates an intuitive natural language interface
between humans and the model~\cite{wu2022promptchainer}.

\begin{comment}
Considering the importance of in-context prompts quality, various studies, therefore, focus on \emph{prompt engineering}~\cite{lu2021fantastically, betz2021thinking, liu2021makes}. Strategies like progressive generation (\ie multi-round text expansion)~\cite{tan2020progressive} and meta-prompting (\ie asking the model to elaborate on the problem)~\cite{betz2021thinking, reynolds2021prompt} attempt to seed LLMs to generate more effective prompts before solving the task.

\end{comment}

In contrast, \emph{few-shot} prompts show the LLM what pattern to follow by feeding it examples of the desired input and output data.
As can be seen in Section~\ref{sec:intro}, given examples on \exinline{Christmas} and \exinline{Halloween}, the LLM would produce a reasonable date for \exinline{Independence Day}.
These examples usually follow consistent structures with meaningful prefixes (\exinline{Holiday: [name] => Date: [date]}), which helps re-emphasize the desired intent~\cite{vaswani2017attention}.
The quality of few-shot prompts heavily relies on the five to thirty in-context examples that demonstrate the intended behavior~\cite{rubin2021learning, lu2021fantastically}, and LLMs can only perform in-context learning if it has seen the corresponding distribution or concept~\cite{min2022rethinking, xie2021explanation, rubin2021learning}. 
If developers omit corner cases in the few examples they created, the task quality can easily be affected~\cite{liu2022wanli}.
For example, without a negative example where we denote ineligible inputs with a placeholder output \exinline{N/A} (\exinline{Holiday: yesterday => Date: N/A}), the LLM would attempt to produce the most plausible ``label'' even for negative examples --- It may try to normalize
\exinline{yesterday} to a most plausible date even though there is no \emph{holiday}.
Our work here tries to help users interactively identify high-quality in-context examples for text transformation.
We review the literature on in-context example selection next.

\subsection{Effective Example Selection}
%In-context learning has been shown to heavily depend on the demonstrative, ``training'' examples, and LLMs can only perform in-context learning if it has seen the corresponding distribution or concept~\cite{min2022rethinking, xie2021explanation, rubin2021learning}. 
Prior work has explored selecting effective demonstrations, and has shown that because pre-trained models possess high-level semantic features, sampling or active learning tends to help identify informative examples~\cite{tamkin2022active}.
In particular, dynamically selecting (retrieving) the most similar demonstrative examples for each given input significantly improves in-context learning performance~\cite{chang2021training, rubin2021learning}.
However, such retrieval methods require fully labeled datasets as the search space.
In contrast, our work studies the scenario where humans craft their personalized in-context functions, and therefore focuses on an unlabeled space.

In the unlabeled search space, prior work has explored effective dataset annotation that can support better in-context learning or few-shot finetuning.
These studies strive to allocate annotation budgets to diverse and representative examples through clustering~\cite{chang2021training} or graph-based search~\cite{su2022selective}.
For example, \citet{su2022selective} built a similarity graph by computing pairwise distances between input sentences and then iteratively selected and annotated examples based on graph density.
They show such selection substantially reduces the annotation cost while achieving high and stable in-context learning performance. 
Despite being effective, these methods sample examples purely for input diversity.
Because our work focuses more on supporting users' interactive function construction, we additionally emphasize \emph{current function quality} in sampling, which helps users track their progress and prioritize improving the current in-context function.
Moreover, these prior studies measures diversity with cosine similarities on input sentence embedding~\cite{reimers-2019-sentence-bert} which, as we argue in Section~\ref{subsec:sampling}, is not reflective of various tasks~\cite{rubin2021learning}.
As a workaround, our work focuses on measuring similarities only on the key phrase embeddings, which leads to more intuitive clusters. 

On the interactive example selection side, our work is perhaps more similar to some literature in programming-by-demonstration (PBD).
For example, \citet{zhang2020interactive} explored effectively selecting examples that can help disambiguate and validate synthesized regular expressions.
We share similar motivations that interactively and iteratively suggest corner cases help synthesize the right function, but unlike PBD where new examples are always \emph{pruning} the function search space, \sysname focuses on \emph{expanding} the function coverage.
Therefore, it is essential to select examples that incentivize people to provide feedback.

\paragraphBold{Active Learning}
Our work is also similar to the aforementioned, effective annotation work~\cite{chang2021training, su2022selective} in the sense that its selection method is akin to sampling approaches in active learning~\cite{settles.tr09, tran2019bayesian}.
The key idea behind active learning is that machine learning models can achieve higher performance with fewer training examples, if it is allowed to choose its own, most informative training examples.
Given a budget, an active learner iteratively selects examples-to-annotate from an unlabeled pool according to some ranking mechanism.
While the previous work is more similar to diversity sampling~\cite{sener2017active}, ours is closer to uncertainty sampling~\cite{lewis1994sequential}, where an active learner queries the instances about which it is least certain how to label.
Because LLMs are generative in nature and do not have clear probabilistic distributions across all ``labels'' as in \eg classification tasks, we estimate uncertainty using the LLM output stability (unanimity voting) which also conveniently serves as a correctness estimation.
This voting strategy is also quite relevant to Query-By-Committee~\cite{seung1992query} where a list of ``committee'' models trained on the same labeled set vote on the labelings of query candidates.
Other work has also been considered directly representing LLM confidence with the average log probability of the entire output~\cite{wang2021want, su2022selective}, an alternative worth comparing against in the future.

Importantly, while many empirical results suggest that active learning is effective, it does suffer from certain limitations. For example, the labeled examples are not drawn \emph{i.i.d} from the underlying data distribution~\cite{settles.tr09}, and therefore can sometimes be imbalanced~\cite{pop2018deep} or less effective than random sampling~\cite{imberg2020optimal}.
Our method will likely share the same limitations, though we leave it to future work to articulate scenarios where \sysname is most useful. \looseness=-1

\subsection{Model-assisted Annotation}

\sysname can also be seen as offering assistance in data annotation (for context learning).
The idea of annotating data with both humans and AI models in the loop has been explored broadly.
In this setup, AIs can play various roles~\cite{yang2017mastering}, \eg they may generate more examples that mimic difficult patterns~\cite{liu2022wanli, ribeiro2022adaptive}, select uncertain examples for people to inspect~\cite{wang2021want}, etc.
\sysname is closer to work encouraging annotators to
find model-fooling examples (``adversarial data collection.'')~\cite{bartolo2021models, dua2019drop, dinan2019build, kiela2021dynabench}.
In particular, \citet{bartolo2020beat} found that in question-answering tasks, models trained on these adversarially collected data can generalize better to more challenging examples.
However, because of the overhead of re-training, their analyses were performed \emph{post-hoc}, \ie they only updated the model offline after collecting a large batch of challenging examples.
In contrast, we leverage the advantage of in-context learning, and directly study the dynamic of in-context function update.

The iterative nature also links \sysname to earlier work in interactive machine learning (IML)~\cite{amershi2014power, wu2019local}. 
IML is a typical paradigm that facilitates iterative and exploratory model understanding and update --- a system explains to users how the current model makes predictions, and users in turn give feedback back to the model, starting the cycle again. 
Labeling is one classic type of IML feedback~\cite{simard2014ice, heimerl2012visual}.
However, because traditional ML tends to focus much more on the surface features (\eg count trigrams in a training example without caring its semantic meanings), users find labeling to be not powerful enough, and prefer richer controls like feature selection~\cite{amershi2014power, patel2008investigating, stumpf2009interacting}.
Since LLMs have some capability to generalize individual examples more broadly to its semantically similar ones, we believe labeling in in-context learning would be more effective, and we use \sysname to reactivate labeling-based IML for in-context learning.

\section{Conclusion}
In this work, we present \sysname, an interactive system for building high-quality demonstration sets for in-context learning. 
\sysname helps users find informative input examples in the unlabeled data, annotate them efficiently with the help of the current version of the learned in-context function, and estimate the quality of said function.
Results from both a simulation study and a 10-person evaluation show \sysname improves in-context function performance, as well as annotator's awareness and handling of diverse patterns. 
Our findings highlight the importance of data slicing and sampling, iterative human-model interaction, and collaborative human-model labeling, and point to interesting future directions such as AI-assisted task definition refinement, more concrete quality metrics that convey the in-context function progress, etc.\looseness=-1

\begin{acks}
This material is based upon work supported by NSF awards 1901386 and 2040196, ONR grant N00014-21-1-2707, and a gift from the Allen Institute for Artificial Intelligence (AI2). 
The authors thank the user study participants for their valuable feedback, and anonymous reviewers for helpful discussions and comments.\looseness=-1
\end{acks}

\bibliographystyle{ACM-Reference-Format}
\bibliography{ref}

%\appendix
%\clearpage\newpage
%\input{appendices/clusters.tex}
%\onecolumn

\end{document}